\documentclass[a4paper]{jpconf}
\usepackage[english]{babel}
\usepackage{amssymb}
\usepackage{graphicx}
\usepackage{epsf}
\usepackage{amsmath}
\usepackage{nicefrac}
\usepackage{esint}
\usepackage{cite}

\def\beq{\begin{equation}}
\def\eeq{\end{equation}}
\def\bea{\begin{eqnarray}}
\def\ea{\end{eqnarray}}
\def\beqa{\begin{equation}\begin{array}{l}}
\def\eeqa{\end{array}\end{equation}}
\def\eeqlab#1{\label{eq:#1}}
\def\figlab#1{\label{fig:#1}}

\def\eref#1{(\ref{eq:#1})}
\def\eeqref#1{Eq.~(\ref{eq:#1})}

\def\Figref#1{Fig.~\ref{fig:#1}}

\def\barr{\left(\begin{array}{c}}
\def\earr{\end{array}\right)}
\def\bmat{\left(\begin{array}{cc}}
\def\emat{\end{array}\right)}
\def\al{\alpha}
\def\be{\beta}
\def\ga{\gamma} \def\Ga{{\it\Gamma}}
\def\de{\delta} \def\De{\Delta}
\def\e{\varepsilon}  

\def\la{\lambda} \def\La{{\Lambda}}

\def\si{\sigma} 
\def\th{\theta}

\def\dd{{\rm d}}
\def\pa{\partial}

\def\pa{\partial}
\def\ra{\rightarrow}
\def\nn{\nonumber}

\def\mathscr{\mathcal}

\def\re{\mbox{Re}}
\def\im{\mbox{Im}}
\def\3d{3-D}

\begin{document}
\title{Sum rules for hadronic contributions to low-energy light-by-light scattering}

\author{Vladimir Pascalutsa, Vladyslav Pauk,  Marc Vanderhaeghen}

\address{Institut f\"ur Kernphysik, Johannes Gutenberg Universit\"at, Mainz D-55099, Germany}

\ead{vladipas@kph.uni-mainz.de}

\begin{abstract}
We present a set of sum rules relating the low-energy light-by-light scattering
to integrals of $\gamma\gamma$ fusion cross-sections and use them to study the hadronic contributions.
\end{abstract}

\section{Introduction}

Some decades ago a general analysis of the forward Compton scattering amplitude allowed Baldin \cite{Baldin:1960},
Gerasimov \cite{Gerasimov:1965et}, Drell and Hearn \cite{Drell:1966jv}
to establish first sum rules expressing the static electromagnetic
properties of the nucleon in terms of its total photoabsorption cross sections.
The Baldin sum rule relates the sum of the electric $\al_E$ and magnetic $\be_M$ polarizabilities 
to an integral of the unpolarized photoabsorbtion cross section $\si$:
\beq
\alpha_E+\beta_M=\frac1 {2\pi^2}\int_0^\infty \! \frac{\dd\nu}{\nu^2}
\, \si(\nu),
\eeq
where $\nu$ is the photon energy in the lab frame.
The Gerasimov-Drell-Hearn (GDH) sum rule  relates the anomalous magnetic moment $\kappa$  of a spin-$1/2$ target
to an integral of the helicity-difference
photoabsorption cross section $\De\si=\si_{3/2}-\si_{1/2}$
(here subscripts stand for the value of total helicity):
\beq
\eeqlab{gSLAC-PUB-0187}
\frac{e^2}{2M^2}\,\kappa^2=\frac1\pi\int_0^\infty \!\frac{\dd\nu}{\nu}
\,\De\si(\nu),
\eeq
with $e$ the charge and $M$ the mass of the target.

A few decades later it has been realized \cite{Roy:1974fz,Gerasimov:1973ja,Brodsky:1995fj} that the GDH sum rule applies to a  photon target, in which case the anomalous magnetic
moment is zero by Furry's theorem, and one simply has:
 \beq
\eeqlab{gammagSLAC-PUB-0187}
0=\frac1\pi\int_0^\infty \!\frac{\dd \nu}{\nu}
\,\Big[\si_2(\nu) - \si_0(\nu)\Big],
\eeq
where $\si_\la$ is the $\ga\ga$-fusion cross section with the total 
helicity $\la$, and $\nu$ is the photon energy in collider kinematics. 

More recently, a systematic derivation of sum rules for light-light system
has been done \cite{Pascalutsa:2010sj}, resulting in sum rules for
the low-energy constants of the Euler-Heisenberg Lagrangian. 
Some details of the sum rules for light-by-light (LbL) scattering will be given below,
with emphasis on perturbative verifications of the sum rules in field theory. 
The case of charged massive spin-1 field is especially interesting.

In the concluding section we shall discuss the way the sum rules can be used
to evaluate the hadronic contributions to the low-energy interaction of light with light.

\section{Sum rules for light-by-light scattering}

The sum rules in question are based on very general
properties of the S-matrix, namely Lorentz and crossing symmetries, analyticity, unitarity, as well the gauge symmetry of the electromagnetic interaction.
To demonstrate this we highlight here the main derivation steps for the
case of $\ga \ga$ systems. 

Denoting LbL scattering ($\gamma \gamma \to \gamma \gamma$) corresponding Feynman and helicity amplitudes as, respectively, $\mathcal{M}$ and $M$, their relation is:
\bea
\eeqlab{genexp}
M_{\la_1\la_2\la_3\la_4} &=& \e^{\ast\mu_4}_{\la_4}(\vec{q}_4) \, \e^{\ast\mu_3}_{\la_3} (\vec{q}_3)\,
\e^{\mu_2}_{\la_2}(\vec{q}_2)\,  \e^{\mu_1}_{\la_1}(\vec{q}_1) \nn\\
& & \times \,  \mathcal{M}_{\mu_1\mu_2\mu_3\mu_4} ,
\ea
where $\e(\vec{q})$ are the photon polarization 4-vectors, $\la$'s are the helicities; for
real photons traveling along the $z$ axis, i.e.\ $\vec{q} = (0,0,\nu)$, the polarization vectors are
$\e_{\la}(\pm \vec{q}) = 2^{-\nicefrac12} ( 0, \mp \la, -i , 0)$.
The Mandelstam variables are defined as
$s=(q_1+q_2)^2 = 4 \nu^2$, $t=(q_1-q_3)^2$, $u=(q_1-q_4)^2$, with $q_i$ the
photon 4-momenta.

In the forward kinematics, where $q_3=q_1$, $q_4=q_2$,
and hence $t=0$, $u=-s$, the general Lorentz structure of the Feynman amplitude is given by:
\bea
\mathcal{M}_{\mu_1\mu_2\mu_3\mu_4} & =&  A(s) \, g_{\mu_4\mu_2} g_{\mu_3\mu_1}
+B(s)\, g_{\mu_4\mu_1} g_{\mu_3\mu_2} \nn\\
&  +&  C(s) \, g_{\mu_4\mu_3} g_{\mu_2\mu_1} \,,
\ea
where $g_{\mu\nu}$ is the Minkowski metric. Crossing symmetry (under $1 \leftrightarrow 3$, or
$2 \leftrightarrow 4$) means in this case
\bea
\mathcal{M}_{\mu_1\mu_2\mu_3\mu_4} & =&  A(u) \, g_{\mu_4\mu_2} g_{\mu_3\mu_1}
+B(u)\, g_{\mu_4\mu_3} g_{\mu_1\mu_2} \nn\\
&  +&  C(u) \, g_{\mu_1\mu_4} g_{\mu_3\mu_2} \,,
\ea
hence $A(-s) = A(s)$, $B(-s) = C(s)$.
 As a result there are three independent nonvanishing helicity amplitudes:
 \bea
  M_{++++}(s) &=& A(s) + C(s),\nn\\
  M_{+-+-}(s) &=& A(s) + B(s),\\
  M_{++--}(s) &=& B(s) + C(s),\nn
 \ea
 satisfying the  following crossing relations:
$ M_{+-+-}(s) =   M_{++++}(-s)$, and
$ M_{++--}(s) = M_{++--}(-s)$.

The principle of (micro-)causality implies that the above functions
are analytic functions of $s$ everywhere in the complex $s$ plane  except along the real axis. For the amplitudes,
 \begin{subequations}
 \bea
f^{(\pm)}(s) &=&  M_{++++}(s)  \pm M_{+-+-}(s) ,\\
g(s) &=&  M_{++--}(s),
\ea
 \end{subequations}
the analyticity  infers the following dispersion relations:
\beq
\re\, \left\{\begin{array}{l} f^{(\pm)} (s)\\
g (s)\end{array} \right\}  = \frac{1}{\pi} \fint\limits_{-\infty}^{\infty} \! \frac{\dd s' }{s'-s} \,
 \im\, \left\{\begin{array}{l} f^{(\pm)} (s')\\
g (s')\end{array} \right\}\,,
\eeq
where $ \fint$ indicates the principal-value integration. These relations hold as long as
the integral converges, and otherwise subtractions are needed.
Because $f^{(\pm)}(-s)=\pm \,f^{(\pm)}(s)$ and $g(-s)=g(s)$,
we can express the right-hand side as an integral over positive $s$
only:
 \begin{subequations}
 \bea
\re\, \left\{\begin{array}{l} f^{(+)} (s)\\
g (s)\end{array} \right\} & = & \frac{2}{\pi}  \fint\limits_{0}^{\infty} \frac{\dd s' \, s'}{s^{\prime 2}-s^2}\,
 \im\, \left\{\begin{array}{l} f^{(+)} (s')\\
g (s')\end{array} \right\}\,,\\
\re\, f^{(-)} (s) & = & -\frac{2s}{\pi} \fint\limits_{0}^{\infty} \dd s' \, \frac{  \im\, f^{(-)} (s')}{s^{\prime 2}-s^2}\,.
\ea
\end{subequations}
In the physical region ($s\geq 0$),  the optical theorem relates
the imaginary part of these amplitudes to the total absorption cross-sections with
definite polarization of the initial $\gamma \gamma$ state:
 \begin{subequations}
\bea
\im\, f^{(\pm)}(s) &=&  - \frac{s}{8} \,[ \,\si_0(s)  \pm \si_2(s) \,] ,\\
\im\, g(s) &=&  - \frac{s}{8} \,[ \,\si_{||}(s)  - \si_\perp (s) \,] .
\ea
\end{subequations}
Substituting these expressions in the above dispersion relations one obtains:
 \begin{subequations}
 \eeqlab{srules}
 \bea
\re\, f^{(+)} (s) & = & -\frac{1}{2\pi}\fint\limits_{0}^{\infty} \dd s'\, s^{\prime 2}\,\frac{\si(s')}{s^{\prime 2}-s^2}\,,
\eeqlab{srulesa}\\
\re\, f^{(-)} (s) & = & -\frac{s}{4\pi} \fint\limits_{0}^{\infty} \dd s' \,\frac{ s'  \, \De\si (s')}{s^{\prime 2}-s^2}\,,\eeqlab{srulesb}\\
\re\, g (s) & = & -\frac{1}{4\pi} \fint\limits_{0}^{\infty} \dd s' \,s^{\prime 2}  \,\frac{
\si_{||}(s')  - \si_\perp (s')}{s^{\prime 2}-s^2}\,,
\eeqlab{srulesc}
\ea
\end{subequations}
where $\si=(\si_0+\si_2)/2=(\si_{||} + \si_\perp)/2$ is the unpolarized
total cross section, and $\De\si=\si_2-\si_0$ (0 or 2 show the total helicity of the circularly polarized
photons, while $\|$ or $\bot$ show if the linear photon polarizations
are parallel or perpendicular).

We next recall that gauge invariance and discrete symmetries constrain the low-energy
photon-photon interaction to the Euler-Heisenberg form \cite{Heisenberg:1935qt}, given by the following Lagrangian density:
\beq
\eeqlab{physics/0605038lagr}
\mathcal{L}_{\mathrm{EH}} = c_1 (F_{\mu\nu}F^{\mu\nu})^2 + c_2 (F_{\mu\nu}\tilde F^{\mu\nu})^2,
\eeq
where $F_{\mu\nu} = \pa_\mu A_\nu - \pa_\nu A_\mu$, $\tilde F^{\mu\nu} =
\e^{\mu\nu\al\be} \pa_\al A_\be$. Expanding the left-hand side and right-hand side of \eeqref{srules} in powers of $s$
and matching them at each order yields a number of sum rules. At 0th order in $s$
we would find
 \bea
0 & = &  \int\limits_{0}^{\infty} \dd s \, \Big[ \si_{||} (s) \pm \si_\perp(s) \Big]\,,
\ea
which cannot work for ``+"  since the unpolarized cross-section is a positive-definite quantity.
Empirically $\si$ shows a slowly rising behavior at large $s$ and the integral diverges.
The assumption of an unsubtracted dispersion relation is violated in this case.
For the ``$-$" case the sum rule is broken too, cf.~\cite{Gerasimov:1973ja} and references
therein.

 At the  first and second orders we find, respectively:
\begin{subequations}
 \eeqlab{s0rules}
 \bea
0 & = &  \int\limits_{0}^{\infty} \dd s\,  \frac{ \De\si (s)}{s}\,,
\eeqlab{s0rulesb} \\
c_1\pm c_2 & = & \frac{1}{8\pi }\int\limits_{0}^{\infty} \dd s\,  \frac{ \si_{||} (s) \pm \si_\perp(s)}{s^2}\,.
 \eeqlab{s0rulesc}
\ea
\end{subequations}
The first sum rule here is the analog of the GDH sum rule mentioned above, 
while the sum rules for the low-energy constants are unique to the $\ga\ga$ system.
Note that according to these sum rules the constants $c_1$ and $c_2$ are positive
definite, in contrast to some previous predictions in the literature
\cite{Kruglov:2001dp}. 

Most of the above arguments will equally hold for the
space-like virtual photons ($q_1^2  < 0 $, $q_2^2 < 0 $), if  written
in a variable which reflects under crossing, e.g.\  $\nu = s-q_1^2-q_2^2$.
The EH Lagrangian must however be extended by terms containing $\pa_\mu F^{\mu\nu}$. For the case when at least one is real such terms are absent
and the above sum rules hold with
only a single modification: $s \to \nu = s- q^2$, where $q^2$ is the 
other photon virtuality.

\section{Verifications in perturbative QED}

\begin{figure}[t]
\centerline{\epsfclipon  \epsfxsize=8,2cm%
  \epsffile{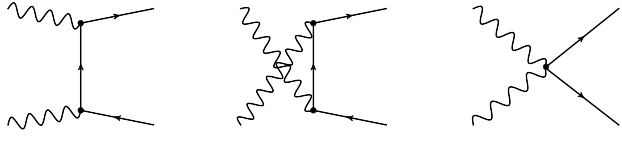}
}
\caption{Pair production in scalar QED.
}
\figlab{scalar}
\end{figure}
The leading-order cross-sections of $\ga\ga$-fusion in QED are 
given by tree-level pair-production diagrams (\Figref{scalar}), 
which can be easily computed
and substituted into the right-hand-side of the sum rules in Eq.~\eref{s0rules}.
In QED of a scalar and a spinor (Dirac) particle the superconvergence sum
rule has thus been verified to leading order the fine-structure constant $\al$.
For example the spin-1/2 pair-production helicity-difference cross section in $\ga^\ast\ga$
collision is given by 
\bea
 && \De \si^{(\ga^\ast\ga\to f \bar f)} (s,q^2,0) =  \frac{8\pi \al^2}{(s-q^2)^2} \, \th(s-4m^2)\\
 &&\, \times  \left\{
  -(3 s+q^2)\sqrt{1-\frac{4m^2}{s}}  + 2 (s+q^2)\, \mathrm{arctanh} \sqrt{1-\frac{4m^2}{s}}
  \, \right\}, \nn
 \ea
  and is plotted in \Figref{xsect} for three different values
  of $q^2$. One sees that in all cases the low- and high-energy contributions cancel.
The fact that
 \beq
  \int\limits_{4 m^2}^{\infty} \dd s\,  \frac{ \De\si^{(\ga^\ast\ga\to f \bar f)} (s,q^2,0)}{s-q^2}=0\,
\eeq
is easily verified for any $q^2 < 4 m^2$.

 \begin{figure}[t]
\centerline{\epsfclipon  \epsfxsize=8 cm%
  \epsffile{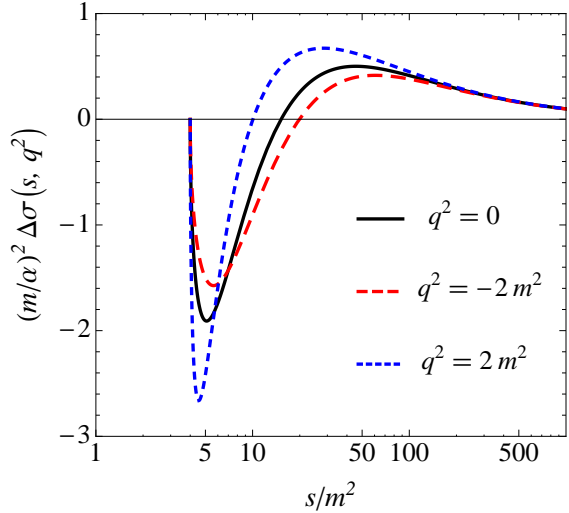}
}
\caption{Helicity-difference cross section of  $\ga^\ast\ga\to f \bar f$ in  QED
 at leading order,
for different photon virtualities.
}
\figlab{xsect}
\end{figure}

Substituting the corresponding linearly-polarized cross section into the sum
rule for EH low-energy constants, we obtain: $c_1=\frac{7\alpha^2}{1440 m^4}$,  $c_2=\frac{\alpha^2}{1440 m^4}$ for the scalar case;   
$c_1=\frac{\al^2}{90 m^4}$, $c_2=\frac{7\al^2}{360 m^4}$ for the spinor case.
This result agrees with the explicit one-loop calculations of low-energy light-by-light
scattering, see e.g.~\cite{Karplus:1950zza,Karplus:1950zz}.

The case of spin-1 QED, describing a charged and massive vector particle, is not as simple. In this case interaction with electromagnetic field can be described in terms of three independent structure functions. Considering the minimal and linear non-minimal couplings we start with the following Lagrangian density:
\beq
\begin{split}
\eeqlab{lagr1}
\mathcal{L}_1=&-\frac 1 4 F_{\mu\nu}F^{\mu\nu}-\frac 1 2 W^*_{\mu\nu}W^{\mu\nu}-M^2W_\mu^*W^\mu+ieW_{\mu\nu}^*A^\mu W^\nu-ieA_\mu W_\nu^*W^{\mu\nu}+e^2A^2W_\mu^*W^\mu+\\
+&iel_1W_\mu^*W_\nu F^{\mu\nu}+{el_2}[(D^*_\mu W^*_\nu)W^\al\partial_\al F^{\mu\nu}+W^*_\al(D_\mu W_\nu)\partial^\al F^{\mu\nu}]/{(2M^2)}, 
\end{split}
\eeq
where $A_\mu$ and $W_\mu$ denote the electromagnetic and vector-boson fields respectively;  $F_{\mu\nu}=D_\mu F_\nu-D_\nu F_\mu$ and $W_{\mu\nu}=D_\mu W_\nu-D_\nu W_\mu$ are the corresponding field-strength tensors, with
$D_\mu=\partial_\mu-ieA_\mu$ the covariant derivative, $e$ the charge and $M$ the mass of the vector boson.
The parameters $l_1$ and $l_2$ contribute to the magnetic and quadrupole moments as: 
\beq
\mu=(1+l_1)\frac{e}{2M}\qquad \mathrm{and} \qquad Q=(l_2-l_1)\frac{e}{M^2}.
\eeq

Computing the cross-section for the tree-level $\ga\ga \to W^+ W^-$ process
in this theory we find that the integral on the right-hand-side of the first (super-convergence)
sum rule in Eq.~\eref{s0rules} diverges, unless $l_1=1$ and $l_2=0$. Only for the
latter choice of the parameters the integral converges and is equal to 0, as it should.
This choice of parameters is realized in the Standard Model, and gives rise to the
so-called "natural values" of the electromagnetic moments.

Computing for the linearly polarized cross sections (for $l_1=1$ and $l_2=0$) 
and using the second sum rule we obtain:
\beq
c_1=\frac{29\alpha^2}{160M^4}\;,\qquad\qquad c_2=\frac{27\alpha^2}{160M^4}.
\eeq
This is in agreement with the one-loop electroweak correction to LbL scattering
(\Figref{bands}),
which we obtained by the low-energy expansion of the expressions of
Bohm and Schuster \cite{Bohm:1994sf}. It is interesting that to obtain
this result in electroweak theory one needs to take care of the Higgs mechanism
as well as the ghosts, while on the side of the sum rule the calculation is much simpler:
tree-level production of massive vector bosons, and a dispersion integral. 

 \begin{figure}[t]
\centerline{\epsfclipon  \epsfxsize=10cm%
  \epsffile{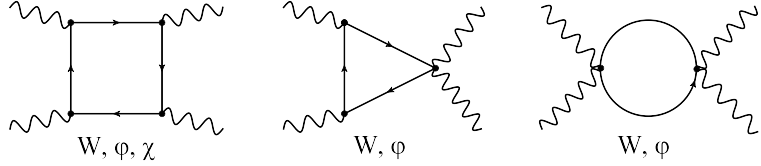}
}
\caption{One-loop Feynman graphs in the standard electroweak theory. Here $W$ stands for the gauge bosons, $\phi$ - Higgs fields and $\chi$ denotes Fadeev-Popov ghosts.
}
\figlab{bands}
\end{figure}

\section{Hadronic contributions to the LbL scattering}

While the role of the SRs in QED becomes fairly clear in these perturbative calculations,
in quantum chromodynamics (QCD), in its non-perturbative regime, it is far less obvious.
We can gain some
insight by looking at  individual contributions to the $\gamma \gamma\to hadrons$  cross sections.

\subsection{Superconvergence sum rule}
The high-energy behavior of $\Delta \sigma$ is determined by a $t$-channel
exchange of unnatural parity and is expected from Regge theory -- in the absence of fixed pole singularities -- to drop as $1/s$ or faster~\cite{Budnev:1971sz}. We therefore expect the
sum rule \eref{s0rulesb} to converge.
The dominant features of the $\gamma \gamma$ to multihadron production  comes firstly from the
Born terms in the $\pi^+ \pi^-$ (or $K^+ K^-$) channels, which each separately obeys the sum rule.
The largest contributions in the hadronic sector are thus expected to come from the resonance production:
$\gamma \gamma \to M$, with $M$ being a meson. It is highly nontrivial to see how the
sum rule is saturated in this case.

As two real photons do not
couple to a $J^P$ = $1^-$ or $1^+$ state due to the Landau-Yang theorem, one expects the
dominant contribution to come from scalar, pseudoscalar, and tensor mesons.
One can express the $\gamma \gamma \to M$ cross section for a meson with spin $J$, mass $m_M$, and total width $\Gamma_{tot}$, using a Breit-Wigner parametrization, in terms of the
decay width $\Gamma_{\gamma \gamma}^{(\Lambda)}$ of the meson
into two photons of total helicity $\Lambda = 0, 2$, as
\begin{eqnarray}
\sigma^{\gamma \gamma \to M}_\Lambda (s) &=& (2 J + 1) \, 16 \pi
 \frac{\Gamma_{\gamma \gamma}^{ (\Lambda)} \, \Gamma_{tot}}{(s - m_M^2)^2 + \Gamma_{tot}^2 m_M^2} \nonumber \\
&\approx& (2 J + 1) 16 \pi^2
\, \frac{\Gamma_{\gamma \gamma }^{(\Lambda)} }{m_M} \, \delta(s - m_M^2),
\label{eq:mesgaga}
\end{eqnarray}
where the last line is obtained in the narrow resonance approximation.
For the pseudoscalar mesons, which can only contribute to the helicity-zero cross section, the narrow resonance approximation is very accurate and allows to quantify their contribution as shown
in Table~\ref{table_ps}. For the pion, this value is entirely driven by the chiral anomaly, which allows
the expression of the $\pi^0$ contribution to the sum rule as $- \alpha^2 / (4 \pi f_\pi^2)$, with $f_\pi = 92.4$~MeV the pion decay constant.

To compensate the large negative contribution to the sum rule from pseudoscalar mesons,
one needs to have an equivalent strength  in the helicity-two cross section, $\sigma_2$.
The dominant feature of the helicity-two cross section in the resonance region
arises from the multiplet of tensor mesons $f_2(1270)$, $a_2(1320)$, and $f_2^\prime(1525)$.
Measurements at various $e^+ e^-$ colliders, notably recent high statistics measurements
by the BELLE Collaboration of the $\gamma \gamma$ cross sections to
$\pi^+ \pi^-$~\cite{Mori:2007bu},  $\pi^0 \pi^0$~\cite{Uehara:2008ep}, $\eta \pi^0$~\cite{Balagura:2007dya},
and $K^+ K^-$~\cite{Abe:2003vn} channels have allowed accurate confirmation of their parameters.
As these tensor mesons were also found to be relatively well described by Breit-Wigner resonances,
we use  Eq.~(\ref{eq:mesgaga}) to provide a first estimate of their contribution to the sum rule.  We
show the results in Table~\ref{table_tensor}, both in the narrow width approximation and using a Breit-Wigner shape, assuming that the tensor mesons pre-dominantly contribute to $\sigma_2$,
as is found by the above-mentioned experimental analyses of decay angular distributions.

 Comparing Tables~\ref{table_ps} and \ref{table_tensor}, we can see that the contribution of the lowest isovector tensor meson composed of light quarks, $a_2(1320)$, compensates to around 70 \% the contribution of the $\pi^0$, which is entirely
governed by the chiral anomaly.
For the isoscalar states composed of light quarks,  the cancellation is even more remarkable,
as the sum of  $f_2(1270)$ and $f_2^\prime(1525)$ cancels entirely, within the experimental accuracy, the combined contribution of the $\eta$ and $\eta^\prime$.

\begin{table}[t]
{\centering \begin{tabular}{|c|c|c|c|}
\hline
& $m_M$  & $\Gamma_{\gamma \gamma} $   &  $\int ds\;  \Delta \sigma  / s$  \\
&  [MeV] &  [keV] &  [nb]  \\
\hline
\hline
$\pi^0$ & 134.98   &  $(7.8 \pm 0.6) \times 10^{-3}$  &  $-195.0 \pm 15.0$   \\
\hline
\hline
$\eta$ &  547.85   &  $0.51 \pm 0.03$  &  $-190.7 \pm 11.2 $  \\
$\eta^\prime$ & 957.66   &  $4.30 \pm 0.15$  & $ -301.0 \pm 10.5$   \\
\hline
Sum $\eta, \eta^\prime$ & & & $-492 \pm 22$ \\
\hline
\end{tabular}\par}
\caption{Sum rule contribution of the lightest pseudoscalar mesons (last column).
The experimental values of meson masses~$m_M$ and $2 \gamma$ decay
widths $\Gamma_{\gamma \gamma}$ are  from PDG~\cite{Amsler:2008zzb}.}
\label{table_ps}
\end{table}

\begin{table}[t]
{\centering \begin{tabular}{|c|c|c|c|c|}
\hline
& $m_M$  & $\Gamma_{\gamma \gamma} $   &  $\int ds\;  \Delta \sigma  / s$ & $\int ds\;  \Delta \sigma  / s$  \\
&  &    &  \rm{narrow res. }  & \rm{Breit-Wigner} \\
&  [MeV] &  [keV] &  [nb]  & [nb] \\
\hline
\hline
$a_2 (1320)$ &  $1318.3$   &  $1.00 \pm 0.06$  &  $134 \pm 8$  &  $137 \pm 8$   \\
\hline
\hline
$f_2 (1270)$  & $1275.1 $   &  $3.03 \pm 0.35$  &  $448 \pm 52$  & $479 \pm 56$   \\
$f_2^\prime (1525)$  & $1525 $   &  $0.081 \pm 0.009$  &  $7 \pm 1$  & $7 \pm 1$   \\
\hline
Sum $f_2, f_2^\prime$ & & & $455 \pm 53$ & $486 \pm 57$ \\
\hline
\end{tabular}\par}
\caption{Sum rule contribution of the lowest tensor mesons. We show both results in the narrow resonance approximation (4th column) and using a Breit-Wigner parametrization (last column).
The experimental values of meson masses $m_M$ and $2 \gamma$ decay
widths $\Gamma_{\gamma \gamma}$ are  from PDG~\cite{Amsler:2008zzb}.}
\label{table_tensor}
\end{table}

Besides the tensor mesons, the subdominant resonance contributions to the $\gamma \gamma$ total cross section arise from the scalar mesons
$f_0 / \sigma(600)$, $f_0(980)$, and $a_0(980)$. A reliable estimate of the scalar mesons requires an amplitude analysis of the partial channels, see e.g.~\cite{Pennington:2008xd}. A future study will estimate more precisely the scalar meson helicity-zero contribution to the sum rule, and elaborate on the cancellation between the
tensor mesons and the (pseudo)scalar meson contributions in the sum rule of
\eeqref{s0rulesb}. Interestingly,  when going to the charm sector,
the sum rule also implies a cancellation between the $\eta_c$ meson, whose contributions amounts to about  $-15.5$~nb,  and scalar and tensor $c \bar c$ states.

In the case of non-zero virtuality of one of the photons, the cross-section of the process $\ga^*\ga\ra M$ is defined not by a Breit-Wigner approximation, but rather by a function of $Q^2$, so-called transition form-factor (TFF). Application of the superconvergent sum rule to these processes can give a useful information of the electromagnetic structure of mesons.
Feynman amplitude of the process $\ga^*\ga\ra M$ for the case of pseudoscalar meson has the form:

\beq
T_{\mu\nu}=ie^2\e_{\mu\nu\al\be}q^\al q'\,^\be F_{\pi^0\ga^*\ga}(Q^2),
\eeq
where $Q^2=-\,q'\,^2$ denots virtuality of the $\ga^*$ and $F_{\pi^0\ga^*\ga}$ is the $\ga^*\ga\ra \pi^0$  TFF.

The leading contribution of the single $\pi^0$ production
to the superconvergent sum rule is found as:

\beq
\int_0^\infty \dd s\frac{\si_0(s)}{s+Q^2}=(4\pi\al)^2\pi\left(\frac1 2 F_{\pi^0\ga^*\ga}(Q^2)\right)^2. 
\eeq
The production of a tensor meson contributes both to the helicity-0 and helicity-2 amplitudes. However the contribution of helicity-0 cross-section is usually negligible, while the helicity-2 
contribution is given by
\beq
\int_0^\infty \dd s\frac{\si_2(s)}{s+Q^2}=(4\pi\al)^2\pi\left(F_{a_2\ga^*\ga}^{\La=2}(Q^2)\right)^2+\mathcal{O}\left(\frac1 {Q^2}\right).
\eeq
Thus, assuming that in the isovector channel   the sum rule is saturated
by  $\pi^0$ ($\La = 0$) and $a_2$  ($\La =2 $) mesons, we obtain at low $Q^2$
the following relation between the TFFs:
\beq
F_{a_2\ga^*\ga}^{\La=2}(Q^2)\approx 
\frac{1}{2} F_{\pi^0\ga^*\ga}(Q^2)\, ,
\eeq
which,  as seen from Tables~\ref{table_ps} and \ref{table_tensor}, is at $Q^2=0$ satisfied to an accuracy of better than 70\%.

\subsection{The low-energy-constant sum rules}

The sum rules
\bea
c_1 & = & \frac{1}{8\pi }\int\limits_{0}^{\infty} \dd s\,  \frac{ \si_{||} (s) }
{s^2}\,,\\
c_2 & = & \frac{1}{8\pi }\int\limits_{0}^{\infty} \dd s\,  \frac{ \si_\perp(s)}{s^2}\,,
 \eeqlab{s2rulesc}
\ea
allow one to assess the size of the hadronic contribution to the low-energy LbL scattering.
For example, the pseudo-scalar meson production in $\ga\ga^\ast$ fusion yields:
\bea
\si_{||} & = & 0 , \nn\\
\si_\perp &=& 4\pi^3 \al^2 \left| F_{M\ga\ga^\ast} (Q^2)\right|^2 \, (s+Q^2) \, \de(s-m_M^2),
\ea
which lead to $c_1=0$ and
\beq
c_2 = \frac{2\pi \Ga_{\ga\ga}}{m_M^5}\,.
\eeq 
The corresponding 
numerical results are given in Table~\ref{2_ps}.
\begin{table}[b]
{\centering \begin{tabular}{|c|c|c|}
\hline
& $c_1$  &  $c_2$   $[\, 10^{-4}$ GeV$^{-4}\,]$   \\
\hline
& & \\
$\pi^0$ & 0   &  $10.8  $     \\
$\eta$ &  0   &  $0.7    $  \\
$\eta^\prime$ & 0 & 0.4     \\
& & \\
\hline
\end{tabular}\par}
\caption{Contribution of the light pseudoscalar mesons to low-energy LbL scattering.}
\label{2_ps}
\end{table}
This result, however, does not aid much in the problem of hadronic LbL contributions
to muon anomaly, $(g-2)_\mu$, where the main effect comes from the LbL
scattering at the hadronic scale, see e.g.\ \cite{Melnikov:2006sr}.

\section*{References}
\bibliography{mybibl}

\providecommand{\newblock}{}
\begin{thebibliography}{10}
\expandafter\ifx\csname url\endcsname\relax
  \def\url#1{{\tt #1}}\fi
\expandafter\ifx\csname urlprefix\endcsname\relax\def\urlprefix{URL }\fi
\providecommand{\eprint}[2][]{\url{#2}}

\bibitem{Baldin:1960}
Baldin A~M {\em et~al.\/} 1960 {\em Nucl. Phys.\/} {\bf 18} 310--317

\bibitem{Gerasimov:1965et}
Gerasimov S 1966 {\em Sov.J.Nucl.Phys.\/} {\bf 2} 430--433

\bibitem{Drell:1966jv}
Drell S and Hearn A~C 1966 {\em Phys.Rev.Lett.\/} {\bf 16} 908--911

\bibitem{Roy:1974fz}
Roy P 1974 {\em Phys.Rev.\/} {\bf D9} 2631--2635

\bibitem{Gerasimov:1973ja}
Gerasimov S and Moulin J 1975 {\em Nucl.Phys.\/} {\bf B98} 349

\bibitem{Brodsky:1995fj}
Brodsky S~J and Schmidt I 1995 {\em Phys.Lett.\/} {\bf B351} 344--348
  (\textit{Preprint} \eprint{hep-ph/9502416})

\bibitem{Pascalutsa:2010sj}
Pascalutsa V and Vanderhaeghen M 2010 {\em Phys.Rev.Lett.\/} {\bf 105} 201603
  (\textit{Preprint} \eprint{1008.1088})

\bibitem{Heisenberg:1935qt}
Heisenberg W and Euler H 1936 {\em Z.Phys.\/} {\bf 98} 714--732
  (\textit{Preprint} \eprint{physics/0605038})

\bibitem{Kruglov:2001dp}
Kruglov S 2001 {\em Annals Phys.\/} {\bf 293} 228--239 (\textit{Preprint}
  \eprint{hep-th/0110061})

\bibitem{Karplus:1950zza}
Karplus R and Neuman M 1950 {\em Phys.Rev.\/} {\bf 80} 380--385

\bibitem{Karplus:1950zz}
Karplus R and Neuman M 1951 {\em Phys.Rev.\/} {\bf 83} 776--784

\bibitem{Bohm:1994sf}
Bohm M and Schuster R 1994 {\em Z.Phys.\/} {\bf C63} 219--225

\bibitem{Budnev:1971sz}
Budnev V, Chernyak V and Ginzburg I 1971 {\em Nucl.Phys.\/} {\bf B34} 470--476

\bibitem{Mori:2007bu}
Mori T {\em et~al.\/} (Belle Collaboration) 2007 {\em J.Phys.Soc.Jap.\/} {\bf
  76} 074102 (\textit{Preprint} \eprint{0704.3538})

\bibitem{Uehara:2008ep}
Uehara S {\em et~al.\/} (Belle Collaboration) 2008 {\em Phys.Rev.\/} {\bf D78}
  052004 (\textit{Preprint} \eprint{0805.3387})

\bibitem{Balagura:2007dya}
Balagura V {\em et~al.\/} (Belle Collaboration) 2008 {\em Phys.Rev.\/} {\bf
  D77} 032001 (\textit{Preprint} \eprint{0709.4184})

\bibitem{Abe:2003vn}
Abe K {\em et~al.\/} (Belle Collaboration) 2003 {\em Eur.Phys.J.\/} {\bf C32}
  323--336 (\textit{Preprint} \eprint{hep-ex/0309077})

\bibitem{Amsler:2008zzb}
Amsler C {\em et~al.\/} (Particle Data Group) 2008 {\em Phys.Lett.\/} {\bf
  B667} 1--1340

\bibitem{Pennington:2008xd}
Pennington M, Mori T, Uehara S and Watanabe Y 2008 {\em Eur.Phys.J.\/} {\bf
  C56} 1--16 (\textit{Preprint} \eprint{0803.3389})

\bibitem{Melnikov:2006sr}
Melnikov K and Vainshtein A 2006 {\em {Theory of the muon anomalous magnetic
  moment}\/} vol 216 (Springer)

\end{thebibliography}

\end{document}